\documentclass[10pt,conference]{IEEEtran}
\usepackage{amssymb,amsthm,amsmath,array}
\usepackage{graphicx}
\usepackage[caption=false,font=footnotesize]{subfig}
\usepackage{xspace}
\usepackage[sort&compress, numbers]{natbib}
\usepackage{stmaryrd}
\usepackage{xcolor}
\usepackage{mathtools}
\usepackage{float}
\usepackage{textcomp}
\usepackage{siunitx}

\begin{document}
\title{Performance Validation of Coded Wavefront Sensing for Quantitative Phase Imaging \\of Static and Dynamic Specimens Using \\Digital Holographic Microscopy}

\author{\IEEEauthorblockN{
        Syed Muhammad Kazim\IEEEauthorrefmark{1}, 
        Franziska Strasser\IEEEauthorrefmark{2},
        Mia Kv{\aa}le L{\o}vmo\IEEEauthorrefmark{2}, 
        Andrii Nehrych\IEEEauthorrefmark{1},
        Simon Moser\IEEEauthorrefmark{2}, 
        \\Micha{\l} Ziemczonok\IEEEauthorrefmark{3},
        Wolfgang Heidrich\IEEEauthorrefmark{4}, 
        Ivo Ihrke\IEEEauthorrefmark{1}, and 
        Monika Ritsch-Marte\IEEEauthorrefmark{2}
    }
    \IEEEauthorblockA{
        \IEEEauthorrefmark{1}Zentrum f{\"u}r Sensorsysteme, University of Siegen, Germany. \\ 
        \IEEEauthorrefmark{2}Institute of Biomedical Physics, Medical University of Innsbruck, Austria. \\ 
        \IEEEauthorrefmark{3}Institute of Micromechanics and Photonics, Warsaw University of Technology, Poland. \\ 
        \IEEEauthorrefmark{4}Visual Computing Center, King Abdullah University of Science and Technology, Saudi Arabia.
        }
}

\maketitle
\begin{abstract}
    Coded wavefront sensing (Coded-WFS) is a snapshot quantitative phase imaging (QPI) technique that has been shown to successfully leverage the memory effect to retrieve the phase of biological specimens. In this paper, we perform QPI on static silica beads and dynamic HEK cells using Coded-WFS. The accuracy of the retrieved phase map is validated using digital holographic microscopy (DHM) for the same specimens. We report comparisons of simultaneous bright-field intensity and optical path delay.
\end{abstract}

\section{Introduction}
\label{sec:intro}

Quantitative phase imaging (QPI) methods enable label-free imaging of weakly absorbing specimens which is crucial for studying the morphology and growth of living cells without interference~\cite{Park:2018}. QPI methods estimate the phase delay of light as it passes through the unknown specimen. Snapshot QPI methods offer support for video-rate imaging, making them an exceptional tool for studying dynamic biological systems~\cite{Cuche:2000, Sung:2009, Wang:2019}.

Coded wavefront sensing (Coded-WFS) is a snapshot QPI technique which requires two measurements: (1) a reference image in the absence of the specimen, and (2) an object image once the specimen is inserted in the optical system. In Coded-WFS, a random phase mask is placed a short distance from the sensor, which produces a speckle pattern when it is illuminated~\cite{Wang:2019, Berto:2017, Wang:2017}. Like the Shack-Hartmann wavefront sensor (SHWFS)~\cite{Malacara-Hernandez:15}, the motion of the diffraction pattern is proportional to the gradient of the specimen's phase. However, unlike the SHWFS, the spatial resolution is higher as the motion of the speckle grains in the diffraction pattern is continuous and not limited by the physical size of the lenslet array.

In this paper, we aim to validate the performance of Coded-WFS in~\cite{Wang:2019} by benchmarking its performance using off-axis digital holographic microscopy (DHM)~\cite{Sanchez:2014} for the same specimens. We examine two specimens: a static silica bead with a homogeneous refractive index (RI) and a dynamic HEK cell. We report the intensities and phases of both. 

\section{Coded-WFS}
\label{sec:coded-wfs}

\begin{figure}[t]
\centering
 \includegraphics[height=0.21\textheight, width=0.45\textwidth]{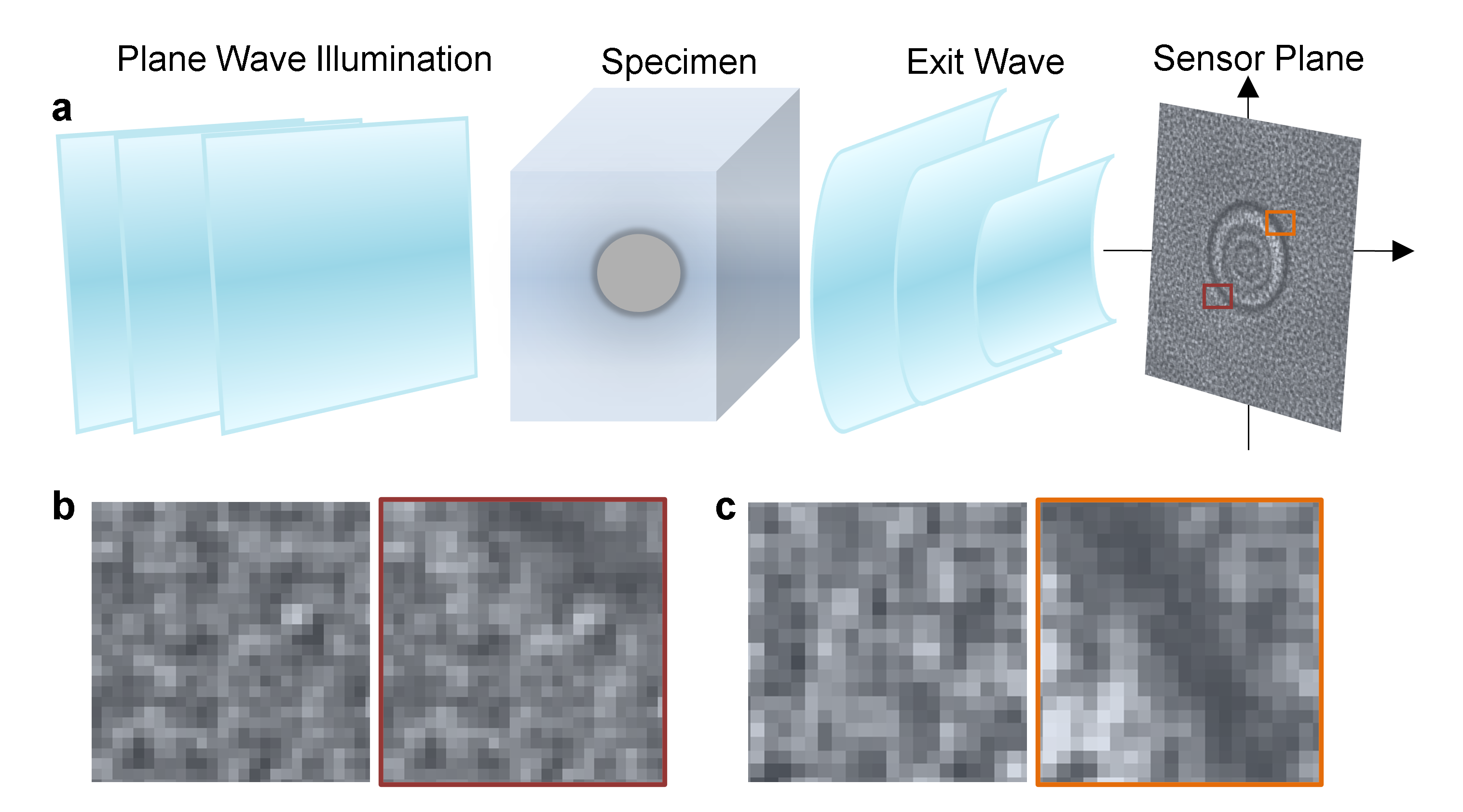}
\caption{\textbf{Measurements using Coded-WFS.} (\textbf{a}) Schematic of Coded-WFS: A plane wave illuminates an immersed specimen and the exit wave is recorded by the Coded-WFS. (\textbf{b}), (\textbf{c}) Two regions in the sensor plane (colored borders) and their corresponding references (without border) illustrate the apparent motion of pixels due to the specimen.}
\label{fig:method}
\end{figure}

In Coded-WFS, a random phase mask is placed close to the image sensor. A pair of reference-object images are recorded, in which the optical system remains identical with the exception that in the reference measurement, the specimen is removed and in the object image, the specimen is inserted in the object plane~\cite{Berto:2017, Wang:2017}. The optical setup is the same as a standard laboratory microscope. Fig.~\ref{fig:method}(a) schematically illustrates the imaging of an immersed sphere where appropriate optics (objective, tube lens) can be placed behind the specimen.

The reference, $I_{0}(r)$, records the speckle pattern of the optical system in the absence of the specimen. When the specimen is inserted in the object plane, the speckle pattern in the image plane is modified, as shown in Figs.~\ref{fig:method}(b) and (c). Coded-WFS leverages the memory effect~\cite{Feng:1988}, which ascribes local changes in the speckle pattern of the recorded image $I(r)$ to the changes in the wave incident on the phase mask. The relation between $I(r)$ and $I_{0}(r)$ is given by,
\begin{equation}
I(r) = I_{0}(r - \frac{z}{k} \nabla \phi (r)),
\end{equation}
where, $z$ is the distance between the phase mask and the camera sensor, $k = 2 \pi /\lambda$ is the wavenumber, $\lambda$ is the illumination wavelength, and $\nabla \phi (r)$ is the gradient of the unknown phase specimen, $e^{j\phi (r)}$. 

The apparent motion between the two measurements is estimated using optical flow algorithms. In this paper, we use the formulation in~\cite{Wang:2019}, which offers simultaneous estimation of the phase and speckle-free brightfield amplitude of weakly absorbing specimens.

\section{Results}

\begin{figure}[th]
\centering
 \includegraphics[height=0.2\textheight, width=0.45\textwidth]{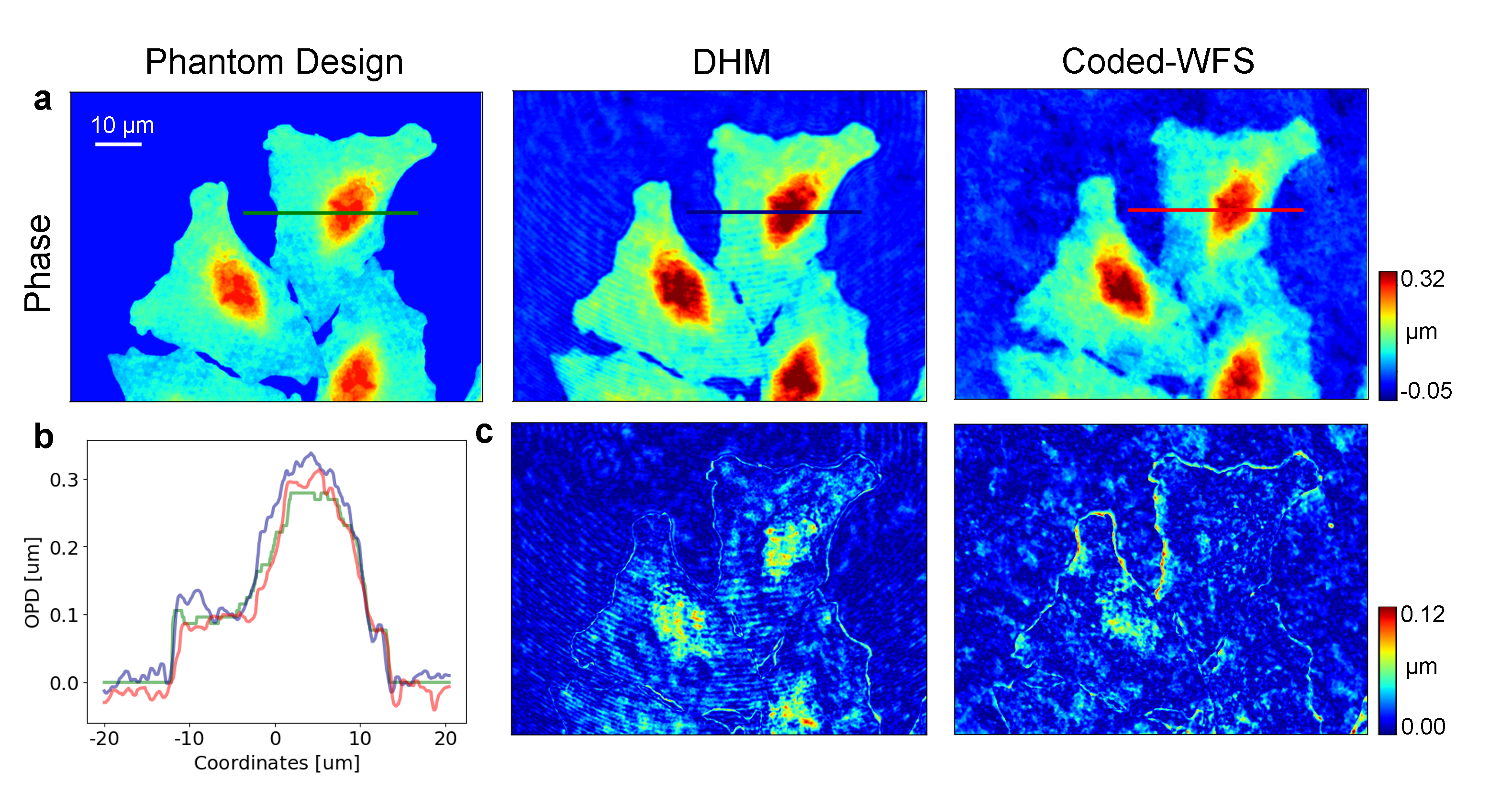}
\caption{\textbf{Performance validation using 3D-printed cluster of artificial HeLa cells.} (\textbf{a}) Designed OPD map of the HeLa cell cluster and measured phase maps (in OPD) of the fabricated phantom using DHM and Coded-WFS. Images were taken under 60x magnification (1.15 and 0.85 NA for DHM and Coded-WFS, respectively). (\textbf{b}) OPDs 
of cross-sections of the phantoms in (a) relative to the immersion. (\textbf{c}) The pixel-wise Euclidean distance between the designed OPD map and the OPDs retrieved using DHM (left) and Coded-WFS (right).}
\label{fig:beads}
\end{figure}

We validate the performance of Coded-WFS and DHM for static phase specimens by imaging an artificial 3D-printed cluster of HeLa cells. Identical models of HeLa cells, fabricated using a polymer with a RI of 1.55 at \SI{633}{\nano\metre}, are placed in different orientations to form a cluster. The designed OPD map of the phantom immersed in Zeiss Immersol 518F (RI=1.518) is shown in Fig.~\ref{fig:beads}(a). The fabricated phantom, which may have errors up to $5 \%$ due to the limited accuracy of the manufactring process, is similarly immersed in Zeiss Immersol before imaging. For Coded-WFS, the specimen was illuminated with a narrowband source (bandwidth of $\approx \SI{5}{\nano\metre}$) centered at \SI{536}{\nano\metre}. Fig.~\ref{fig:beads}(a) shows the retrieved phases of the phantom using DHM and Coded-WFS. Figs.~\ref{fig:beads}(b) and~(c) compare the cross-sections and the pixel-wise Euclidean distance, respectively, of the OPDs retrieved by the two techniques with the phantom's designed OPD map. To visualize the OPD of the phantom relative to the immersion, the mean OPD of the background has been subtracted from the retrieved phase.

\begin{figure}[th]
\centering
 \includegraphics[height=0.325\textheight, width=0.45\textwidth]{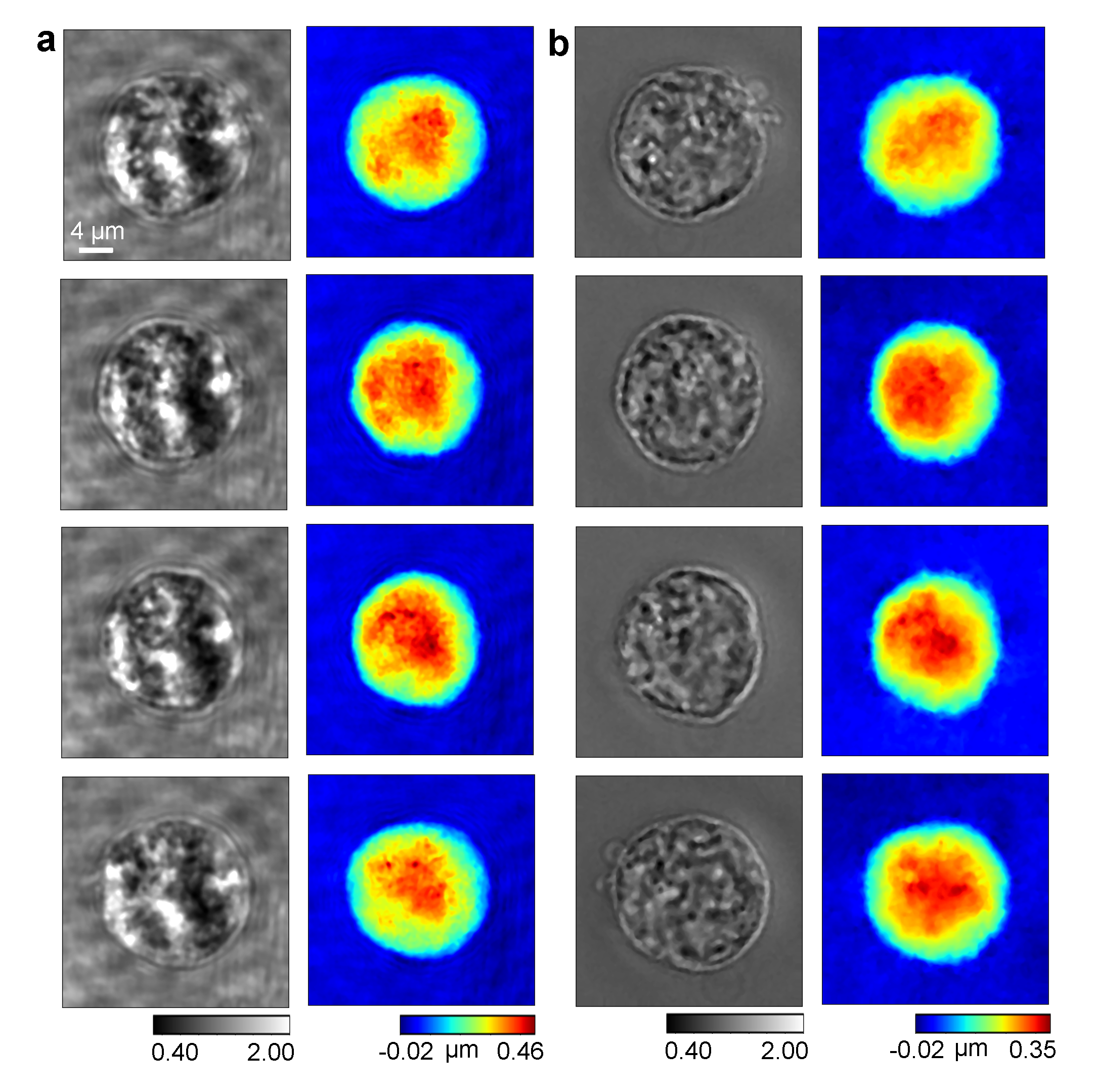}
\caption{\textbf{Video-rate ($\approx \textrm{30~fps}$) QPI.} Quantitative reconstructions of intensity (left) and OPD (right) of corresponding frames using (\textbf{a}) DHM and (\textbf{b}) Coded-WFS of a rotating HEK cell. Images were taken under 60x magnification (1.15 NA).}
\label{fig:superstar}
\end{figure}

In the second experiment, we illustrate that the snapshot QPI capability of Coded-WFS enables QPI of dynamic biological specimens. Here, a single HEK cell is trapped and actuated using an acoustofluidic trapping device described in~\cite{Lovmo:2021}, such that the cell rotates about the axis orthogonal to the imaging axis at $\approx \SI{0.86}{\radian\per\second}$.
Both techniques sequentially record the same rotating specimen at $\approx 30~ \textrm{fps}$, which is only limited by the sensor technology. For Coded-WFS, the cell was illuminated using a white light LED (Thorlabs MWWHL4). Fig.~\ref{fig:superstar} shows agreement between the retrieved intensities and OPDs of the cells by DHM and Coded-WFS in multiple poses.

\section{Conclusion and Future Work}

In this work, we have reported good agreement between the complex fields retrieved by Coded-WFS and DHM by illustrating the intensities and OPDs of static and dynamic specimens. We experimentally demonstrate that DHM provides a higher spatial resolution, provided the sensor's pixel pitch and the objectives are the same. As the Coded-WFS is compatible with narrowband and broadband illumination, the retrieved intensities have reduced diffraction artifacts because the partial coherence acts as an averaging operator.

The video data captured by trapping and actuating the HEK cell can be considered for optical tomography to retrieve the 3D RI distribution of the cell. However, as the rotation is not completely controlled, the pose information of the cell with respect to each frame is not available, preventing a direct application of tomography algorithms.

\section{Acknowledgements}
This work has been supported by the German Research Foundation (DFG) under grant FOR 5336 (IH 114/2-1).

\bibliographystyle{IEEEtran}
\bibliography{references}
\end{document}